\documentclass[a4paper]{jpconf}  % for double-spaced preprint
\usepackage[dvips]{graphicx}
\usepackage{dcolumn}   % needed for some tables
\usepackage{bm}        % for math
\usepackage{amssymb}   % for math
\usepackage[utf8]{inputenc}
\usepackage{mathrsfs}
\usepackage{amsmath}
\usepackage{hyperref}
\usepackage{textcomp}
\usepackage{subfigure}
\usepackage{color}

\begin{document}

\title{Scattering of kinks in a non-polynomial model}
\author{D Bazeia,$^{1}$ E Belendryasova$^{2}$ and V A Gani$^{2,3}$ }
\address{$^1$Departamento de F\'isica, Universidade Federal da Para\'iba, 58051-900 Jo\~ao Pessoa, Para\'iba, Brazil}
\address{$^2$National Research Nuclear University MEPhI (Moscow Engineering Physics Institute), 115409 Moscow, Russia}
\address{$^3$National Research Center Kurchatov Institute, Institute for Theoretical and Experimental Physics, 117218 Moscow, Russia}
\ead{vagani@mephi.ru}

\begin{abstract}
We study a model described by a single real scalar field in the two-dimensional space-time. The model is specified by a potential which is non-polynomial and supports analytical kink-like solutions that are similar to the standard kink-like solutions that appear in the $\varphi^4$ model when it develops spontaneous symmetry breaking. We investigate the kink-antikink scattering problem in the non-polynomial model numerically and highlight some specific features, which are not present in the standard case.
\end{abstract}

\section{Introduction}

Topological defects are extremely important for theoretical physics. They are used for describing models in cosmology high energy physics, quantum and classical field theory, condensed matter and so on. In this context, the $(1+1)$-dimensional models are of special interest, because the dynamics of many three- and two-dimensional systems can be reduced to the one-dimensional models \cite{GaKsKu01}--\cite{GaKu.SuSy.2001}.

Consider a field-theoretical model in $(1+1)$-dimensional space-time with a real scalar field $\varphi(x,t)$ with the Lagrangian density
\begin{equation}\label{eq:Lagrangian}
\mathcal{L}=\frac{1}{2}\left(\frac{\partial\varphi}{\partial t}\right)^2-\frac{1}{2}\left(\frac{\partial\varphi}{\partial x}\right)^2-U(\varphi),
\end{equation}
where the potential $U(\varphi)$ is a non-negative function with two or more degenerate minima (vacua of the model): $\varphi_1^{(0)}$, $\varphi_2^{(0)}$, \dots, $U(\varphi_1^{(0)})=U(\varphi_2^{(0)})=...=0$. The equation of motion for the field $\varphi$ is
\begin{equation}\label{eq:eqmo}
\frac{\partial^2\varphi}{\partial t^2} - \frac{\partial^2\varphi}{\partial x^2} + \frac{dU}{d\varphi}=0.
\end{equation}
The energy functional corresponding to the Lagrangian \eqref{eq:Lagrangian} reads
\begin{equation}\label{eq:energy}
E[\varphi] = \int_{-\infty}^{\infty}\left[\frac{1}{2} \left( \frac{\partial\varphi}{\partial t} \right)^2 + \frac{1}{2} \left( \frac{\partial\varphi}{\partial x} \right) ^2 + U(\varphi)\right]dx.
\end{equation}
In this model, kinks and antikinks are static solutions which minimize energy and interpolate between neighboring vacua.

\section{The sinh-deformed $\varphi^4$ model}

\begin{figure}[h!]
\centering
\begin{minipage}{0.3\linewidth}
\subfigure[\:Potentials]{
\includegraphics[width=\linewidth]{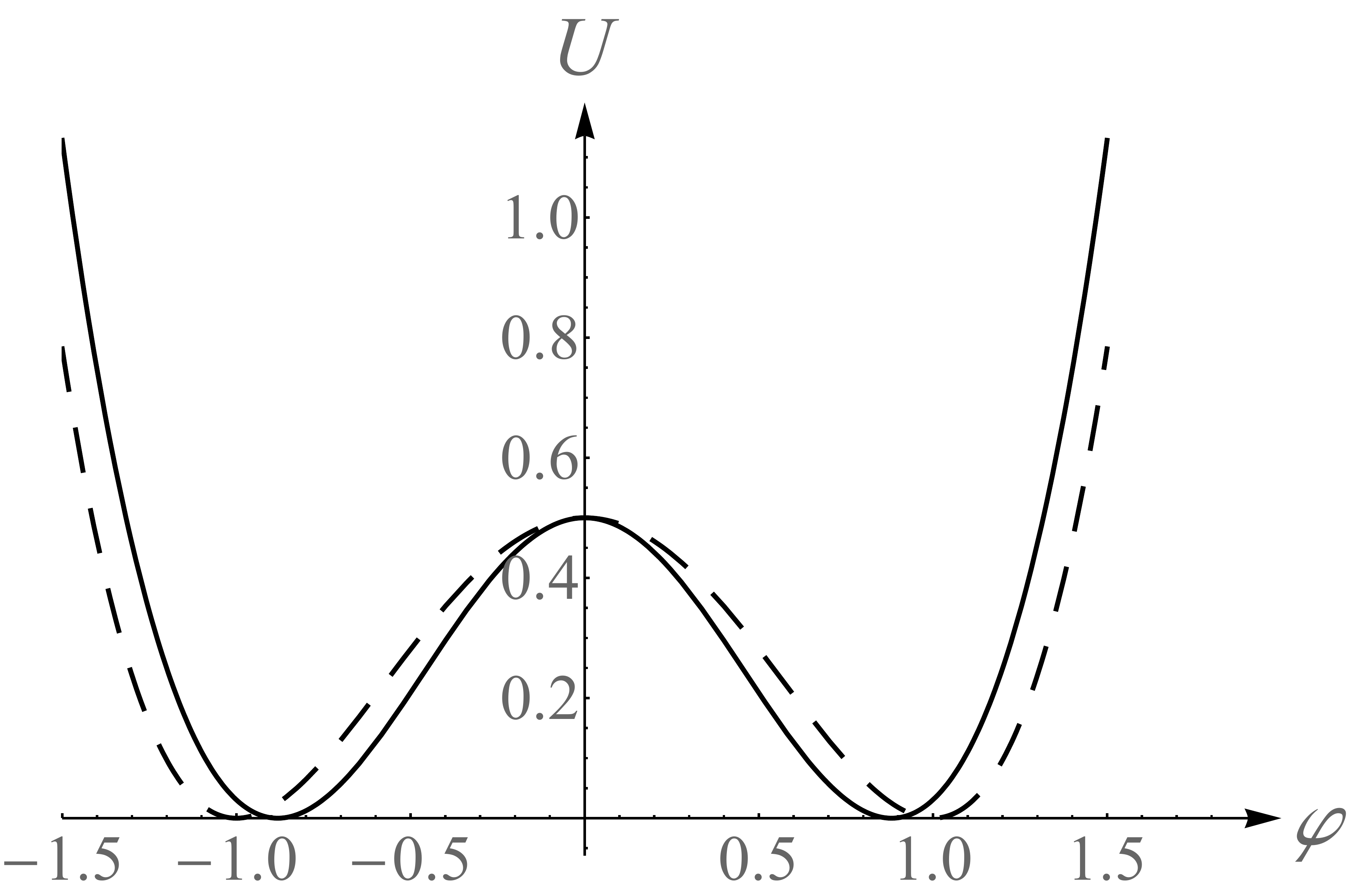}
\label{fig:potentials}
}
\end{minipage}
\hspace{20mm}
\begin{minipage}{0.3\linewidth}
\subfigure[\:Kinks]{
\includegraphics[width=\linewidth]{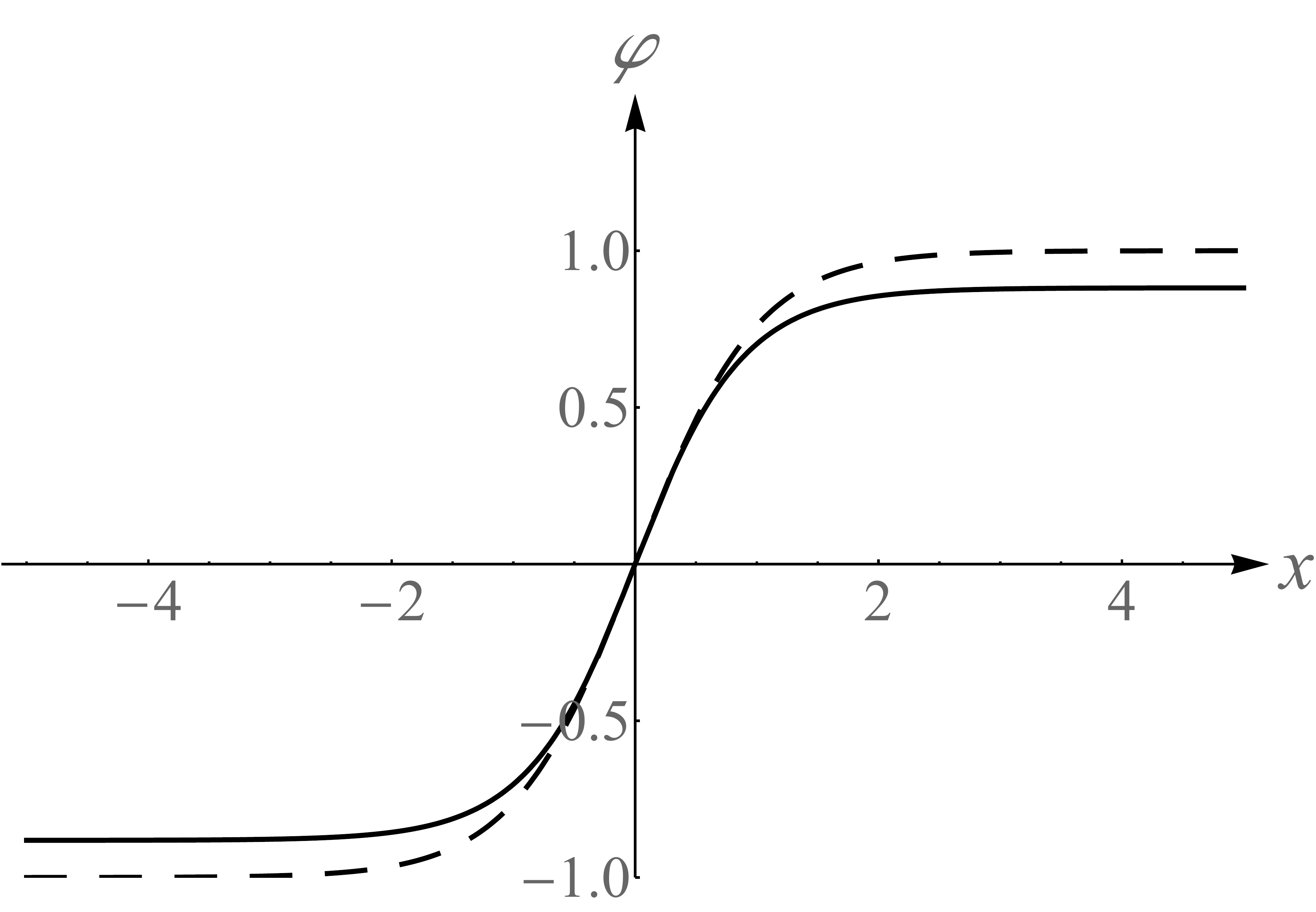}
\label{fig:kinks}
}
\end{minipage}
\caption{Potentials and kinks of the $\varphi^4$ (dashed curves) and the sinh-deformed $\varphi^4$ (solid curves) models.}
\label{fig:potentials_and_kinks}
\end{figure}

This paper deals with the model which we call the sinh-deformed $\varphi^4$ model. The potential $U(\varphi)$ of this model can be obtained from the potential $V(\varphi)=\frac{1}{2}(1-\varphi^2)^2$ of the standard $\varphi^4$ model by deformation procedure which was elaborated and applied in \cite{Bazeia.PRD.2002}--\cite{Bazeia.IJMPA.2017}:
\begin{equation}\label{deform}
U(\varphi) = \frac{V(\varphi\to f(\varphi))}{(df/d\varphi)^2},
\end{equation}
where ``$f(\varphi)$'' --- the deforming function and ``$\varphi\to f(\varphi)$'' means that $\varphi$ is replaced by $f(\varphi)$. 
So,
\begin{equation}
\label{eq:potential_sinh_phi4}
U(\varphi) = \frac{1}{2}\:\mbox{sech}^2\varphi\left(1-\sinh^2\varphi\right)^2.
\end{equation}
This potential has two minima with $\varphi_\pm^{}=\pm\mbox{arsinh}\:1$, $U(\varphi_\pm^{})=0$, figure \ref{fig:potentials}. The kink of the new model, $\varphi_\mathrm{k}^{\mathrm{(new)}}(x)$, is related with kink of the $\varphi^4$ model, $\varphi_\mathrm{k}^{\mathrm{(old)}}(x)$, by the inverse deforming function $f^{-1}$,
\begin{equation}
\varphi_\mathrm{k}^{\mathrm{(new)}}(x) = f^{-1}(\varphi_\mathrm{k}^{\mathrm{(old)}}(x)).
\end{equation}
Using $\varphi_\mathrm{k}^{\mathrm{(old)}}(x)=\tanh x$, we obtain kink and antikink of the sinh-deformed $\varphi^4$ model:
\begin{equation}
\label{eq:sinh_phi4_kinks}
\varphi_\mathrm{k}^{}(x) = \mbox{arsinh}(\tanh x), \quad \varphi_\mathrm{\bar k}^{}(x) = -\mbox{arsinh}(\tanh x),
\end{equation}
see figure \ref{fig:kinks}.

\section{Kink-antikink scattering}

We studied the kink-antikink scattering in the sinh-deformed $\varphi^4$ model using the initial configuration
\begin{equation}\label{eq:in_cond_sinh_phi-4}
\varphi(x,t) = \mbox{arsinh}\left[\tanh\left(\frac{x+x_0^{}-v_\mathrm{in}^{}t }{\sqrt{1-v_\mathrm{in}^2}}\right)\right] - \mbox{arsinh}\left[\tanh\left(\frac{x-x_0^{}+v_\mathrm{in}^{}t}{\sqrt{1-v_\mathrm{in}^2}}\right)\right] - \mbox{arsinh}\:1,
\end{equation}
where $x_0^{}$ is half-distance between two kinks, and they start moving towards each other with the initial velocities $v_\mathrm{in}^{}$. We used $2x_0^{}=10$ and the following parameters of the numerical scheme: $\delta t=0.008$, and $\delta x=0.01$

We found the critical value of the initial velocity $v_\mathrm{cr}^{}\approx 0.4639$, which separates two different regimes of the kinks scattering. At $v_\mathrm{in}^{}>v_\mathrm{cr}^{}$ the kinks escape to infinities after one collision. At the initial velocities $v_\mathrm{in}^{}<v_\mathrm{cr}^{}$ there is the kinks' capture and formation of their bound state, and many escape windows. The picture looks similar to the $\varphi^4$ case.

\begin{figure}[t!]
\centering\subfigure[\:\ A bound state of two oscillons at $v_\mathrm{in}=0.44190$]{
\begin{minipage}{0.45\linewidth}
\includegraphics[width=0.85\linewidth]{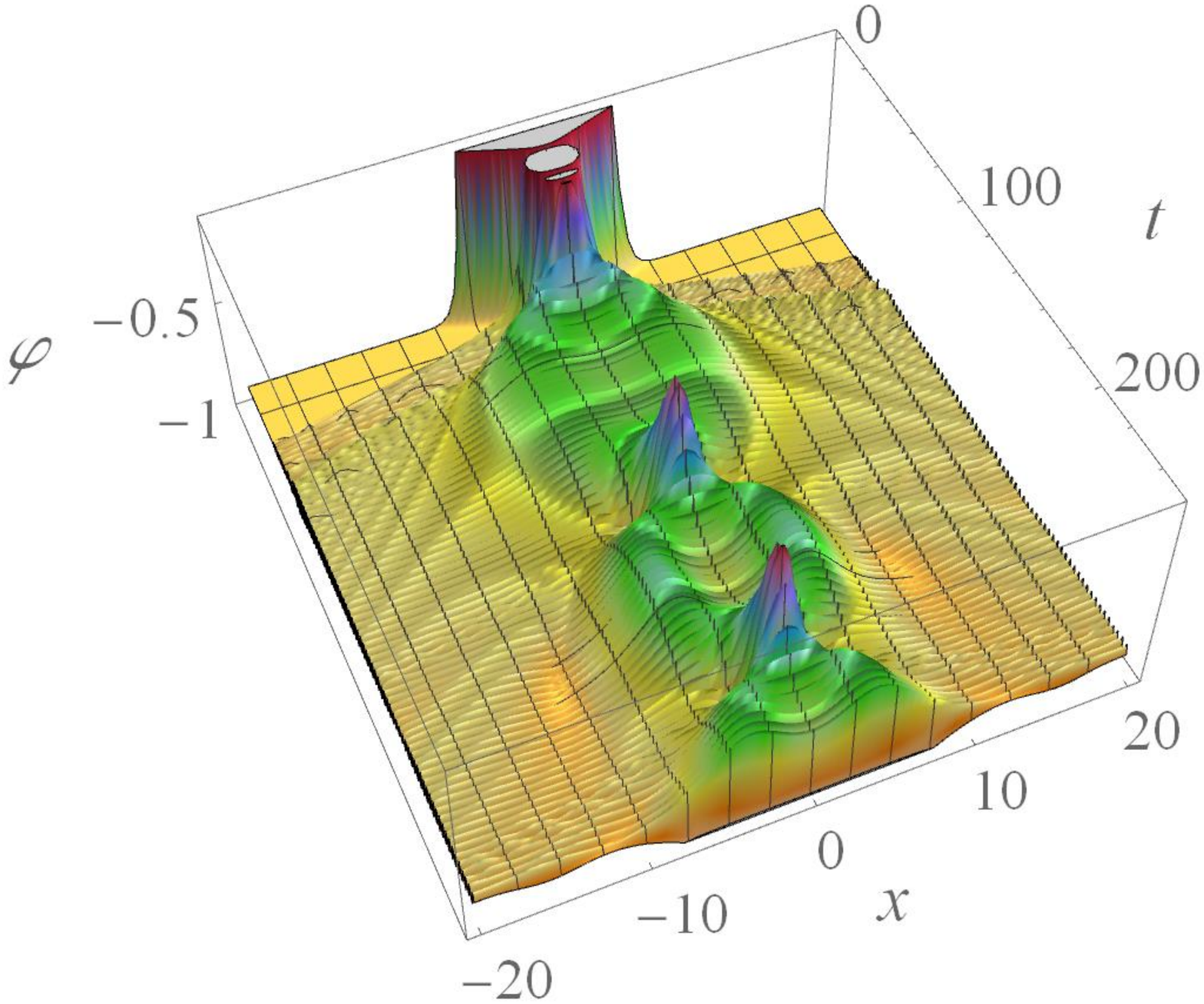}
\end{minipage}
\label{fig:osc_bion}}
\hspace{5mm}
\centering\subfigure[\:\ An escape of two oscillons at $v_\mathrm{in}=0.44195$]{
\begin{minipage}{0.45\linewidth}
\includegraphics[width=0.85\linewidth]{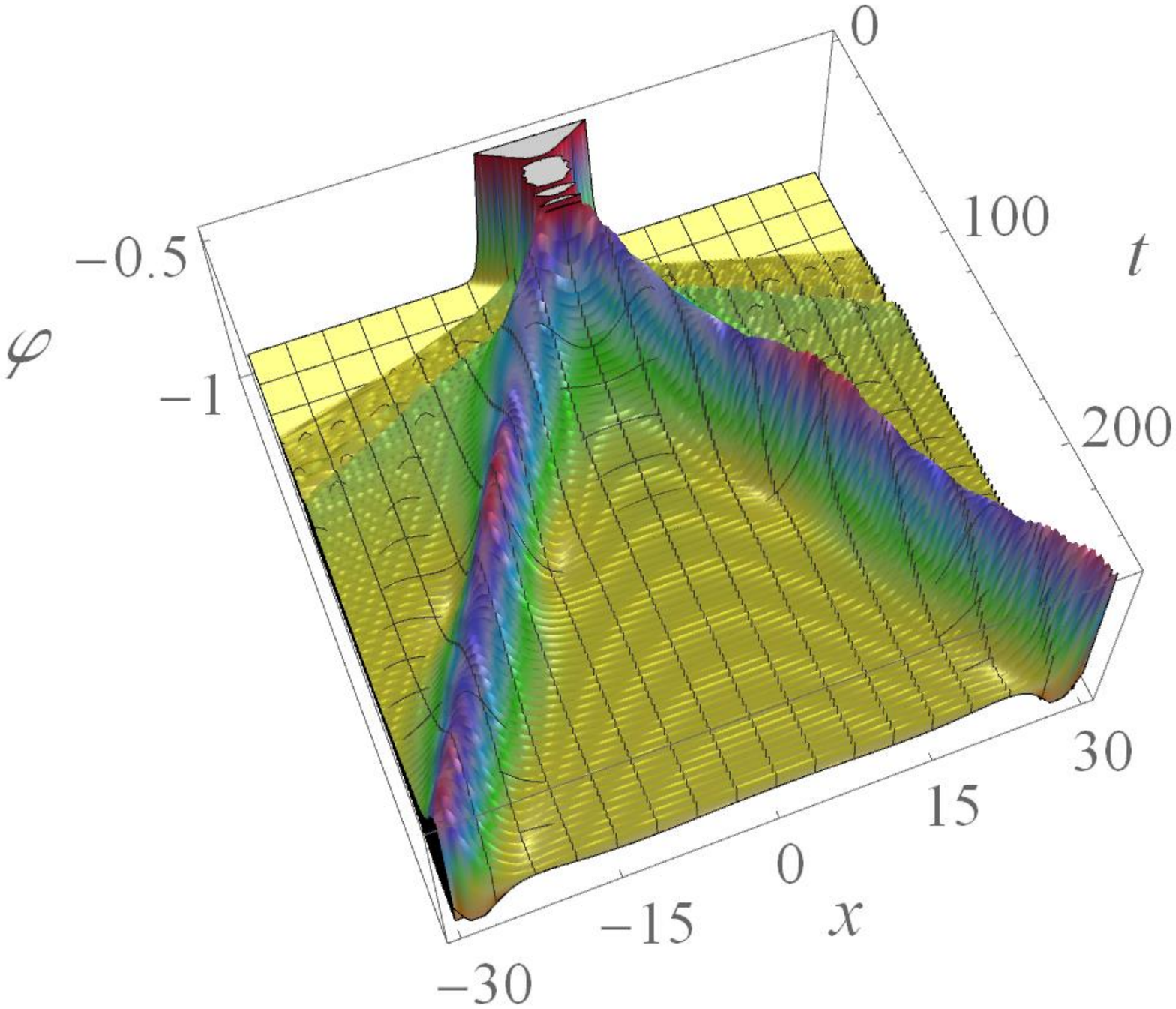}
\end{minipage}
\label{fig:osc_escape}}
\caption{Formation of two oscillons in the kink-antikink collision.}
\label{fig:oscillons}
\end{figure}

\begin{figure}[t!]
\vspace{-30ex}
\centering\includegraphics[width=0.8\linewidth]{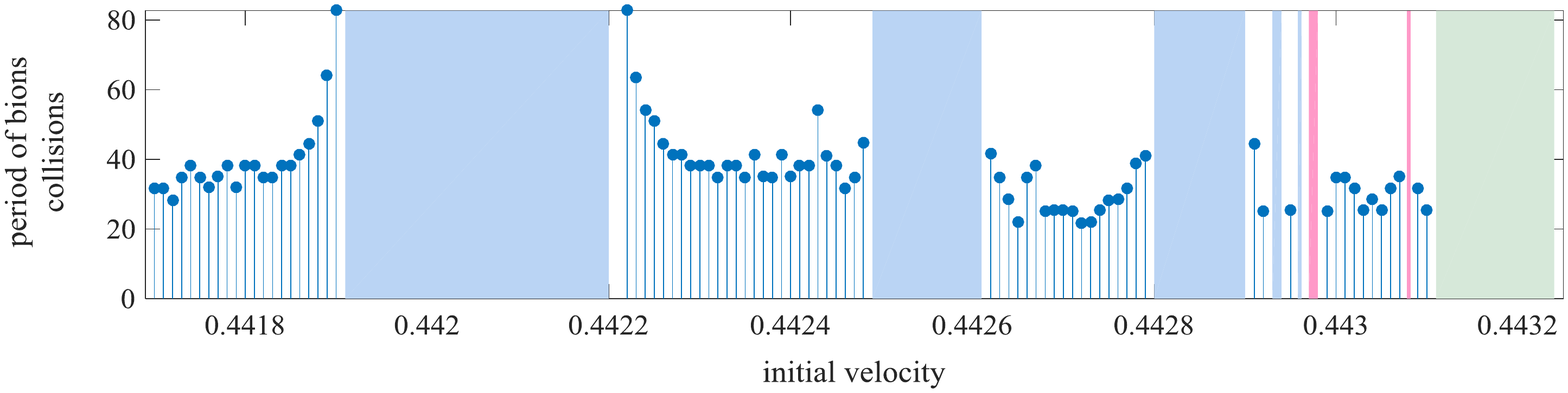}
\caption{The period of oscillations of the oscillons as a function of the initial velocity of the colliding kinks. The blue-shaded areas denote the escape of oscillons. The green-shaded area denotes the 2-bounce escape window for kinks, the pink-shaded areas denote the 4-bounce escape windows for kinks.}
\label{fig:period}
\end{figure}

Besides that, in the range $v_\mathrm{in}^{}<v_\mathrm{cr}^{}$ we have found new interesting result, which has not been observed for the $\varphi^4$ kinks. In many cases the final configuration looked like two oscillons, which can form a bound state, figure \ref{fig:osc_bion}, or escape to infinity, figure \ref{fig:osc_escape}.  In a bound state the oscillons' oscillate near each other with the period which depends on the initial velocity of the colliding kinks, see figure \ref{fig:period}.

\section{Conclusion}

We studied the kink-antikink scattering in the sinh-deformed $\varphi^4$ model.

There is the critical initial velocity $v_\mathrm{cr}^{}\approx 0.4639$, which separates two regimes of the kinks scattering. At $v_\mathrm{in}^{}>v_\mathrm{cr}^{}$ the kinks escape to infinities after one collision. At the initial velocities $v_\mathrm{in}^{}<v_\mathrm{cr}^{}$ we observed formation of kinks' bound state and resonance phenomena. This pattern is similar to that found in the $\varphi^4$ model. Moreover we found new interesting phenomenon --- formation of two oscillons which can form a bound state or escape to infinities. It would be very interesting to investigate the nature of appearance of oscillons. We believe that it could become a subject of further study.

\section{Acknowledgments}

This work was performed using resources of the NRNU MEPhI high-performance computing center. The research was supported by the Brazilian agency CNPq under contracts 455931/2014-3 and 306614/2014-6, and by the MEPhI Academic Excellence Project under contract No.~02.a03.21.0005, 27.08.2013.

\end{document}